# Spatio-spectrally tailored multimode metasurface lasers in the visible range

Ayesheh Bashiri,[*,1,2] Aleksandr Vaskin,[2] Katsuya Tanaka,[1,2,4] Muyi Yang,[1,2,4] Thomas Pertsch,[1,2,3,4] and Isabelle Staude,[1,2,4]

[1]Institute of Solid-State Physics, Abbe Center of Photonics, Friedrich Schiller University, Jena, 07743, Germany.

[2]Institute of Applied Physics, Abbe Center of Photonics, Friedrich Schiller University, Jena, 07745, Germany.

[3]Fraunhofer-Institute of Applied Optics and Precision Engineering IOF, Jena, 07745, Germany.

[4]Max Planck School of Photonics, Jena, 07745, Germany.

*Corresponding author. Email: Ayesheh.bashiri@uni-jena.de

**Abstract**

Spectrally engineered multifrequency nanolasers are highly desirable for on-chip photonics, multiplexed biosensing, and display technologies, yet achieving them within a single compact platform remains challenging. Here, we demonstrate multimode lasing from symmetry-broken TiO$_2$ metasurfaces integrated with an SU8 slab waveguide containing Rhodamine 6G. By co-engineering guided-mode resonances, surface lattice resonances near Rayleigh anomalies, and quasi-bound states in the continuum, we realize complementary high-Q feedback pathways that overlap with the gain spectrum. The direction of the lasing emission is tailored through outcoupling via second-order Bragg diffraction and Rayleigh anomaly conditions, supporting both normal and oblique emission. Experiments reveal discrete lasing outputs across ≈100 nm bandwidth (548–648 nm), spanning nearly the full Rhodamine 6G emission band, with thresholds as low as ∼7 nJ per pulse (35.7 µJ/cm$^2$) and up to four concurrent lasing peaks from a single device. These results establish a metasurface-dye platform for multifrequency and angle-selective lasing, opening new opportunities for compact, multifunctional nanophotonic sources.

**Introduction**

Compact laser sources capable of multiwavelength emission with controllable directionality are highly desirable for next-generation photonic technologies.[1-3] Realizing such functionality within a single planar architecture requires a photonic platform that supports multiple optical resonances with distinct spectral and angular characteristics, strong light confinement, and minimal losses. All-dielectric metasurfaces, composed of periodic arrays of high-index nanostructures, are well suited for this purpose because they can host a rich set of high-quality (Q)-factor resonances in the visible with far lower intrinsic loss than plasmonic counterparts.[4-9,11,26-30] Combined with subwavelength thickness and geometric tunability, dielectric metasurfaces, when integrated with an efficient gain medium, provide an ideal route to spectrally and directionally engineered multifrequency lasing within a compact footprint. Low-threshold operation, however, requires more than simply minimizing loss. The lasing mode must also exhibit strong spatial and spectral overlap with the gain medium to maximize light–matter interaction.[10,11]

In metasurfaces cladded with a gain layer, several physically distinct routes can supply the required optical feedback. A widely used pathway is diffraction-coupled band-edge lasing, in which the lattice couples in-plane propagating modes to the far-field at specific points of the dispersion.[12,13] Such band edges can originate from guided-mode resonances (GMRs),[14,15] in which waveguide-like modes in the gain layer couple to the lattice through Bragg scattering,[16,17] or from surface lattice resonances (SLRs),[18-21,30,31] namely collective modes arising from the diffractive coupling of particle resonances near the Rayleigh anomaly (RA). At the band edge, the group velocity approaches zero, resulting in strong field localization and an enhanced Q-factor. Under the second-order Bragg condition, these modes couple efficiently to the zeroth diffraction order, enabling normal-direction emission, while satisfying the Bragg condition for other diffraction orders allows lasing at defined angles.

Another pathway is through bound states in the continuum (BICs), which are resonant modes lying above the light line yet completely decoupled from radiative channels. In dielectric metasurfaces, BICs typically arise when the mode field distribution possesses a symmetry (e.g., even/odd parity) incompatible with that of a plane wave at normal incidence, thereby preventing out-of-plane coupling. Introducing a controlled asymmetry, such as a geometric perturbation, converts the ideal BIC into a quasi-BIC, where the radiative leakage becomes finite but can be tailored to remain very small. This enables outcoupling in the normal direction while preserving the high Q-factor required for efficient optical feedback.[22-27]

Each of these routes has been leveraged separately for lasing. Azzam et al. reported near-$\Gamma$, BIC-assisted lasing from (titanium dioxide) $TiO_2$ nanoresonator arrays via lattice–dipole coupling, achieving single and multimode lasing at small off-normal angles.[28] Low-threshold lasing was enabled by high-Q, symmetry-protected BICs originating from first- and second-order transverse-electric (TE)-polarized slab-waveguide modes coupled to the periodic array of $TiO_2$ nanoantennas.[11] Barth et al. achieved lasing from a large-area 2D material using a metasurface supporting both quasi-BIC and GMR modes, enabling $\Gamma$-point lasing with similar thresholds[29]. Yang et al. demonstrated low-threshold $\Gamma$-point lasing in hybrid SLR $Si_3N_4$ metasurfaces by deliberately breaking structural symmetry to convert ideal BICs into quasi-BICs.[30] In a related plasmonic structure, Guan et al. realized white-light lasing by sandwiching three different dye solutions between metasurfaces of different periods, using multimodal SLRs to generate RGB lines with tunable relative intensities, underscoring spectral multiplexing within a single device.[31] Despite these advances, co-engineering multiple mode classes within one device, specifically, SLRs, GMRs, and quasi-BICs, to achieve simultaneous multiwavelength and multi-angle lasing remains unexplored.

In this work, we introduce symmetry-broken all-dielectric metasurfaces composed of L-shaped[32,33] TiO$_2$ nanoresonators ($n \approx 2.4$) on a silicon dioxide (SiO$_2$) substrate ($n \approx 1.45$). The metasurfaces are spin-coated with an SU8 layer ($n \approx 1.6$) doped with Rhodamine 6G (Rh6G) laser dye (Figure 1a). TiO$_2$ combines a high refractive index with negligible absorption across the visible spectral range. Rh6G offers a broad gain spectrum, high quantum yield, and ease of integration.[34,35] With air above and SiO$_2$ below, the SU8 core forms an asymmetric slab waveguide, while the metasurface at the SU8-SiO$_2$ boundary plays a dual role: it supports localized multipolar resonances that, under intentional symmetry breaking, evolve into quasi-BICs, while its periodicity provides the reciprocal-lattice momentum needed for SLR formation, GMR coupling, and second-order Bragg feedback at band edges, enabling vertical outcoupling at Γ.

Through careful control of slab thickness, lattice period (P), and asymmetry, this platform brings GMRs, SLRs, and quasi-BICs into deliberate coexistence. Leveraging multiple mode types is essential to achieve broad spectral coverage for lasing within a single dye-integrated metasurface. Simply increasing the waveguide thickness to support higher-order GMRs sacrifices compactness and typically weakens their phase-matched coupling to the lattice and the vertical channel at the Γ-point. SLRs, in contrast, arise only at discrete band edges defined by diffraction orders (Rayleigh conditions), leading to relatively large free spectral ranges for a given period. Quasi-BICs provide high Q-factor resonances, but their fields remain concentrated largely within the high-index resonators, limiting overlap with the dye unless hybridized with lattice or guided modes.

By combining modes of different physical origins, our platform achieves complementary resonances with tailorable spacing and coupling efficiency, enabling multiple spectrally engineered lasing peaks across the Rh6G gain bandwidth. Depending on the design, emission can be directed either along the surface normal or into oblique angles, thus unifying spectral and directional control within a single metasurface, an advance not realized in previous multimode lasing systems.

Experimentally, we demonstrate lasing across the full visible gain bandwidth of Rh6G, spanning from 548 nm (green) to 648 nm (red). Notably, the 548 nm lasing represents the shortest-wavelength laser reported for Rh6G, while the 648 nm peak extends to the dye's longest accessible wavelength. Depending on the structural design, we observe up to three or even four simultaneous lasing peaks, all with relatively low thresholds on the order of ~7 nJ (35.7 µJ/cm$^2$) pump pulse energy (PPE). As an important example, for a suitably optimized metasurface design, lasing peaks are observed at 550, 570, and 630 nm, corresponding to green, yellow, and red emission, respectively.

Our approach demonstrates passive spectral and angular multiplexing in a monolithic, all-dielectric platform, without the need for active tuning or multiple gain materials. This work establishes a new paradigm for multifunctional nanophotonic sources with potential impact on next-generation optical communication, sensing, and display technologies.

## Results and discussion

For the design of the multifrequency metasurface lasers, we started by selecting P so that the first-order RAs ($\lambda_{RA} = n_{eff} \times P$) fell within the Rh6G gain bandwidth (Figure 1c), yielding an SLR near $\lambda$ =550 nm from the SiO$_2$ substrate and another near $\lambda$ =600 nm from the SU8 cladding. Next, using finite-element simulations (COMSOL Multiphysics), we designed a square lattice of cubic nanoresonators with a small notch in the corner to break in-plane symmetry and generate high-Q quasi-BICs within the spectral gain bandwidth.

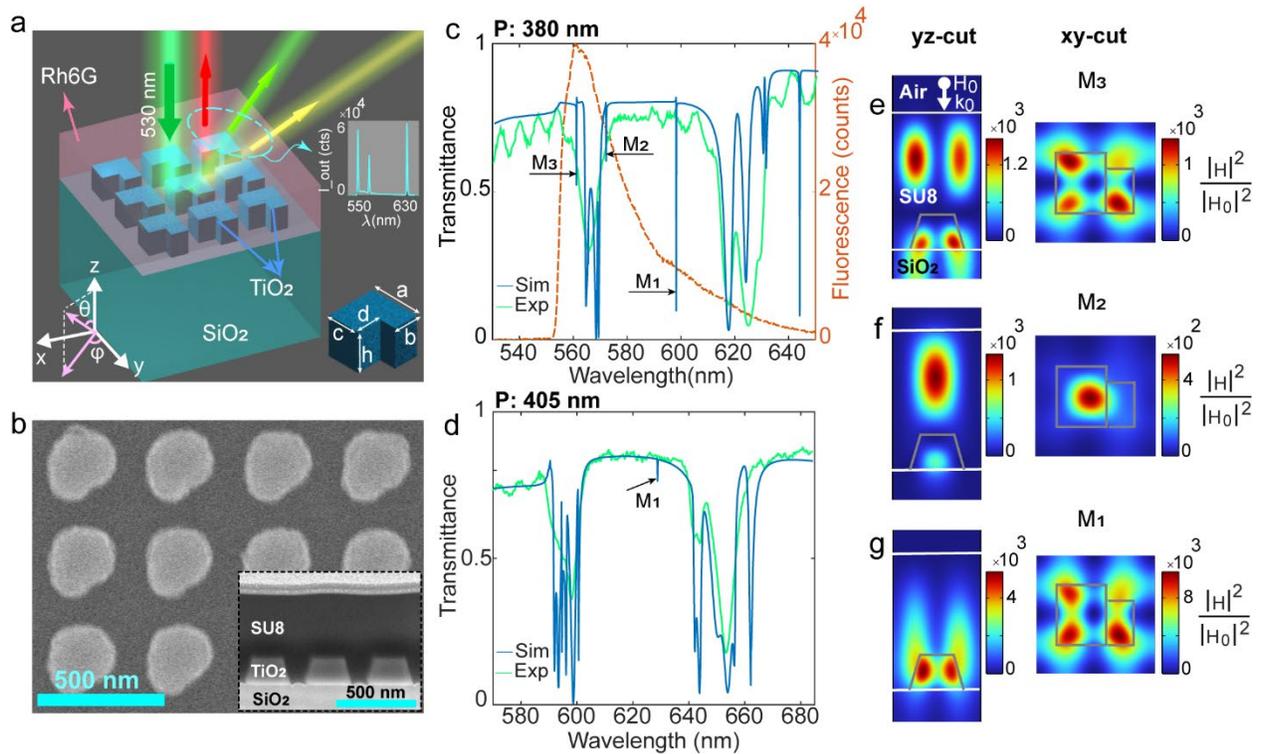

"**Figure 1**. (a) Schematic illustration of lasing metasurface showing an array of L-shaped TiO$_2$ nanoresonators on a SiO$_2$ substrate, integrated with an Rh6G-doped SU8. Upon a 532 nm excitation, the coupled system generates triple-mode lasing at 550 nm (green), 570 nm (yellow), and 630 nm (red). (b) Top-view scanning-electron micrograph (SEM) of the fabricated metasurface with a period of 380 nm. The inset shows a focused-ion-beam cross-section SEM of the metasurface coated with Rh6G-doped SU8 of 570 nm thickness. (c, d) Measured and numerically calculated linear optical transmission spectra of the coated metasurfaces with a period of (c) 380 nm (LMS1) and (d) 405 nm (LMS2), for y-polarized normal incidence illumination. The high-Q modes leading to lasing at normal incidence for LMS1 are denoted with $M_1$–$M_3$. The red dashed line in (c) represents the fluorescence spectrum of Rh6G. (e-g) Calculated near-field intensity profiles of LMS1 for the modes $M_1$–$M_3$ in the cross-section through the center of the nanoresonator for y-polarized normal incidence illumination. All mode profiles are normalized with respect to the intensity of the incident plane wave. The nanoresonator outline is indicated with gray solid lines."

We calculated normal-incidence transmission under x- and y-polarized illumination (section S1). We performed parametric sweeps to adjust the SU8 thickness in order to tune the resonances of the lowest-order GMRs to fall within the Rh6G gain bandwidth, while the resonator dimensions and notch depth were varied to achieve the same for the quasi-BIC mode. The resulting spectra for y-polarized incident light and for lasing metasurfaces with P=380 nm (LMS1) and 405 nm (LMS2) are shown in Figure 1c,d, respectively. Corresponding spectra for x-polarized incident light are shown in Figure S1. The geometrical parameters according to the inset of Figure 1a listed in Table 1. For LMS1, three sharp high-Q modes (Q-factors in SI), labeled as $M_1$ at 597 nm, $M_2$ at 571 nm, and $M_3$ at 560 nm, fall within the Rh6G gain bandwidth. In LMS2, all resonances redshift. In particular, $M_1$ appears at 630 nm.

We repeated the simulation for an otherwise identical, notch-free structure, where we observed that all three resonances disappear (Figure S2), confirming that they are intrinsically dark modes that emerge as high-Q quasi-BICs only through symmetry breaking. We then calculated near-field intensity cross-sections through the resonator center for $M_1$–$M_3$ (Figure 1e–g) and also performed a multipole decomposition (section S4). The results show $M_1$ (600 nm) is dominated by an in-plane magnetic quadrupole (MQ), which is intrinsically symmetry-protected. The fields are confined inside the TiO$_2$ nanoresonators and extending into SU8, consistent with SLR-type coupling near the SU8-side RA. The yz-cut shows a single-lobed $E_z$-dominant distribution (Figure 1g, S4c) resembling a transverse-magnetic (TM)$_0$-like GMR, though the guided contribution is weaker than the lattice coupling.[18] $M_2$ (570 nm) is dominated by an out-of-plane magnetic dipole (MD$_z$), likewise dark in the symmetric case. Its field profiles reveal two vertically separated antinodes with a central node and a dominant $H_z$ component (Figure 1f, S4b), matching the TE$_1$-like GMR in the SU8. Finally, $M_3$ (560 nm) is dominated by the same in-plane MQ mode as $M_1$. The yz-cut shows two antinodes separated by a node along z with a dominant $E_z$ component (Figure 1e, S4a), resembling a TM$_1$-like GMR in the SU8 slab, while spectral proximity to the SiO$_2$-side RA and substrate-extended fields indicate SLR hybridization.[18,36]

Using the optimized design parameters for the nanoresonators and SU8 thickness, we fabricated metasurfaces with periods ranging from 330 to 415 nm using electron beam lithography (EBL) followed by reactive ion etching (RIE) (see supporting information (SI)). To compensate for fabrication tolerances, the lateral dimensions of the nanoresonators were varied by adjusting the exposure dose to ensure accurate realization of the designed structures. Figure 1b shows a scanning electron micrograph (SEM) of LMS1, exhibiting a slight corner rounding; otherwise, the dimensions closely match the design (Table 1).

Table 1. Designed and experimentally realized geometrical parameters for LMS1 and LMS2.

| Parameters* (nm) | | a | b | c | d | h | SU8 height |
|---|---|---|---|---|---|---|---|
| LMS1 (P=380) | Sim | 142 | 119 | 105 | 79 | 153 | 593 |
| | Fab ≈ | 162 | 113 | 102 | 67 | 156 | 570 |
| LMS2 (P=405) | Sim | 175 | 81 | 110 | 75 | 153 | 593 |
| | Fab ≈ | 162 | 113 | 102 | 67 | 156 | 570 |

*See the inset of Figure 1a for the definitions.

Subsequently, the fabricated metasurfaces were spin-coated with a 570 nm-thick SU8 layer doped with Rh6G dye (section S7). A focused-ion-beam (FIB) cross-section of the coated metasurface is shown in Figure 1b, inset. Nanoresonator sidewalls exhibit a slope of ≈ 70° relative to the substrate, as a consequence of the etching process. This feature was incorporated in all design and simulation steps to ensure accurate performance modeling.

To pre-characterize the coated metasurfaces, we measured polarization-resolved, normal-incidence transmission for LMS1 and LMS2. The measured spectra for y-polarized and x-polarized incident light are shown in Figure 1c,d, and S2a, b. A good qualitative agreement with numerically calculated spectra is observed, considering that the numerically predicted high-Q resonances are not resolved due to finite angular collection (NA≈0.04, ±2.5°) and the spectrometer resolution (full-width at half-maximum (FWHM)=1.34 nm). However, these modes become evident under optical pumping as lasing peaks and are further confirmed through far-field emission pattern characterization.

To experimentally characterize the lasing characteristics of the dye-integrated metasurfaces, we performed power-dependent fluorescence spectroscopy. To this end, we pumped them using a 532 nm pulsed laser (0.5 ns pulse duration, 1 Hz repetition rate) over a PPE range of 0–9 nJ (0–45.9 µJ/cm$^2$), with each spectrum integrated over a 1 s acquisition period. The collimated beam passed a square aperture, which was imaged onto the selected metasurface to provide uniform illumination confined to the metasurface patch, with a square spot size of 140 µm per side. Emission was collected with a 0.9-NA objective (polar angle $\theta = \pm 64°$) and directed either to a spectrometer or to a camera imaging the back-focal plane (BFP) of the objective (see SI). Figure 2a presents power-dependent emission from LMS1 with increasing PPE. At 6.7 nJ, the spectrum follows the broad Rh6G fluorescence (spontaneous emission). At 6.9 nJ, a narrow line appears at 600 nm, consistent with $M_1$ in the simulated transmission (Figure 1c). With further increase (≈7.6 and 8.4 nJ), additional peaks emerge at 574 nm ($M_2$) and 559 nm ($M_3$). Figure 2b exhibits output intensity ($I_{out}$) (blue) and FWHM (orange) as functions of PPE. The former exhibits the characteristic S-curve for all three peaks, with clear thresholds of ~6.9–8.4 nJ PPE (35.2–42.9 µJ/cm$^2$), confirming lasing at 600, 574, and 559 nm. Simultaneously, the FWHM narrows down from ≈ 30 nm to ≈ 1.34 nm (limited by spectrometer resolution).

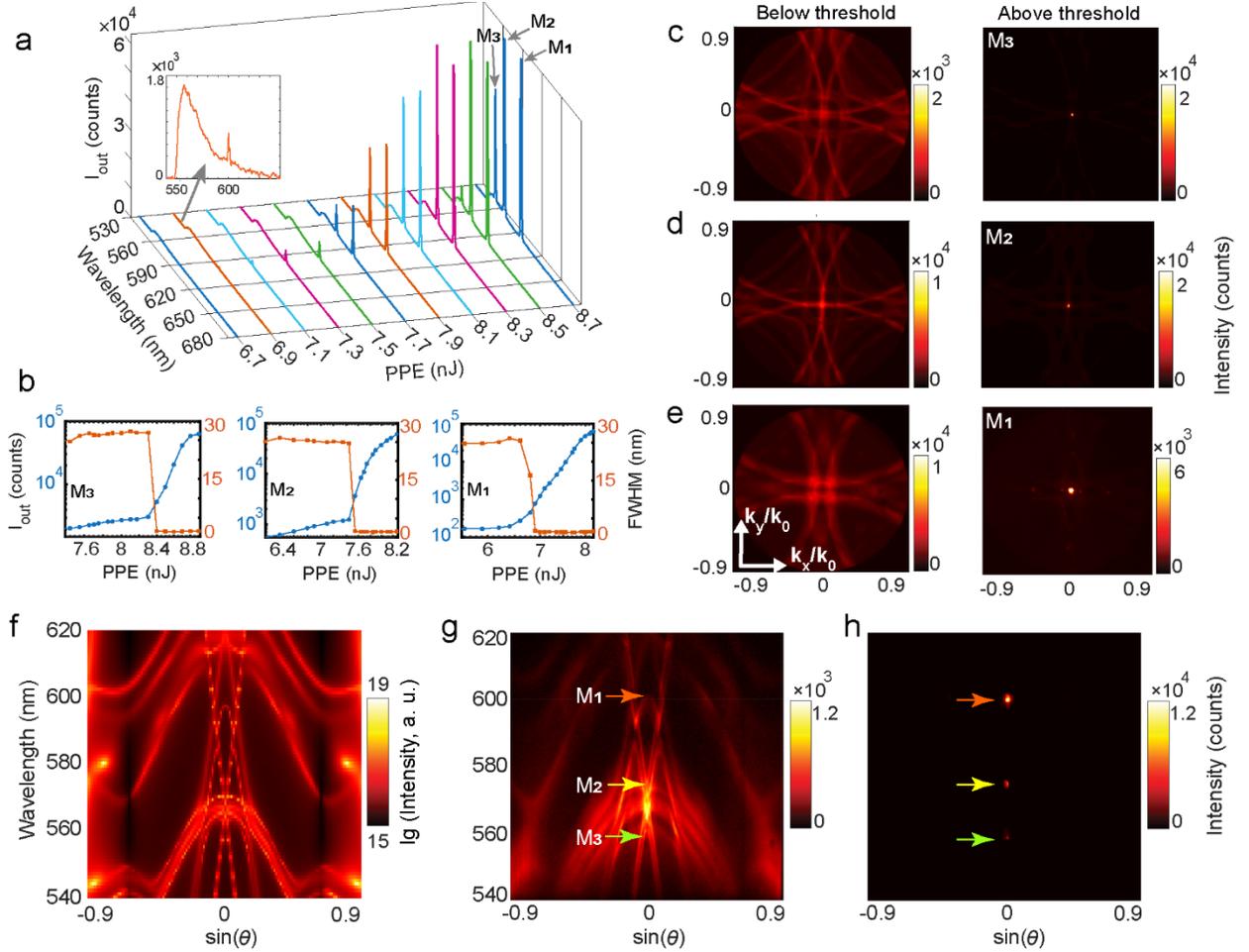

**Figure 2**. (a) Evolution of the emission spectra of LMS1 at different PPEs. The inset shows the initial emergence of the lasing peak associated with resonance $M_1$ at 6.9 nJ PPE. (b) Maximum output intensity $I_{out}$ at different PPEs (S-curve, (log–log scale), blue) and FWHMs (orange) for $M_3$–$M_1$. (c-e) Measured BFP images of LMS1 using BP filters centered at (c) 560 nm ($M_3$), (d) 570 nm ($M_2$), and (e) 600 nm ($M_1$). Below threshold (left) and with a repetition rate of 500 Hz and an integration time of 2 s for improved signal strength, it shows the characteristics of spontaneous emission into the spatially dispersive metasurface modes, dominated by diffractive features. Above threshold (right) at PPE (1 Hz, 1 s) (c) 8.6 nJ, (d) 7.9 nJ, and (e) 6.1 nJ, lasing emission becomes dominant, manifesting as a bright, sharp feature in the center, corresponding to beaming. (f) Numerically calculated angular resolved spectrum for an angle $\varphi = 0$ and $\theta$ up to

±64° averaged over TE and TM polarization of the incident light. (g, h) Measured momentum-resolved spectra below (g) and above (h) threshold."

Polarization-dependent measurements are presented and discussed in section S10. Furthermore, an eigenfrequency calculation with a small artificial gain in the SU8 layer confirmed that modes $M_1$–$M_3$ are the first to acquire net amplification upon increasing the gain, consistent with the experimentally observed lasing peaks (SI).

Next, we characterized the far-field emission of LMS1 using BFP imaging (section S8) with band-pass (BP) filters (10 nm bandwidth) centered at 560, 570, and 600 nm (Figure 2c–e). Note that these images are iso-frequency slices of the angular dispersion, rather than the exact lasing ridge. For example, $M_2$ lases at ≈574 nm; however, the 570 nm filter represents the nearby band edge features. Below threshold (left), the images display the characteristic diffraction circles of the periodic lattice, consistent with spontaneous emission coupled into the lattice modes. Above threshold, the emission at all three wavelengths concentrates into a sharp, focused spot at the center of the corresponding BFP images ($k_x, k_y = 0$), where ($k_x, k_y$) are the in-plane components of the photon momentum, demonstrating normal-direction outcoupling of the lasing emission.

To further characterize the lasing emission, we performed momentum-resolved spectroscopy by spectrally resolving a thin slice of the BFP images around $k_y = 0$ (azimuthal angle $\varphi = 0$). Momentum-resolved emission spectra (Figure 2g) reveal two steep diffractive branches folding at Γ (second-order Bragg condition). The band edges corresponding to the $SiO_2$- and SU8-side RA are observed at ≈570 and 600 nm, respectively. Dispersion flattens at these edges, indicating slow-light regimes, with $M_3$ located at the lower edge and $M_2$ at the upper edge of the 570 nm band, while $M_1$ resides at the higher wavelength (600 nm) band edge. We reproduced the main features of the measured momentum-resolved spectra using simulations based on the reciprocity principle[37] (Figure 2f, SI). Above threshold (Figure 2h), the emission collapses into sharp Γ-centered features, confirming normal-direction, highly collimated lasing (spot in *k*-space) from all three modes.

Next, to investigate lasing characteristics in LMS2 (P=405 nm), we repeat all the measurements starting from power-dependent fluorescence spectroscopy. As shown in Figure 3a, three lines emerge at 550, 570, and 630 nm with close threshold values of ≈7.8 nJ (39.8 µJ/cm$^2$) for the first two and ≈7 nJ (35.7 µJ/cm$^2$) for the latter.

The shorter-wavelength modes, denoted as $M_4$ and $M_5$, as described in the next section, are new resonances, while the 630 nm peak is the red-shifted counterpart of $M_1$ from LMS1. In the inset (8.7 nJ PPE), we can clearly see that the spectrum spans green to red, demonstrating an exceptionally broad band for a single-dye lasing metasurface platform. The S-curves and power-dependent FWHM evolutions (Figure 3b) confirm lasing for all three modes. Figure 3c–e shows BFP images for the three lasing modes $M_5$, $M_4$, and $M_1$, measured with BP filters at 550 and 570 nm and a long-pass (LP) filter at 610 nm, respectively. Below threshold (left), the patterns show the characteristic spontaneous emission of periodic metasurfaces, appearing as broad diffraction rings from the nanoresonator lattice. For $M_5$ (550 nm) and $M_4$ (570 nm), no Γ-point states are present (Figure 3c,d). In contrast, for $M_1$ (630 nm), the band-crossings at Γ remain evident, indicating that the condition for SLR formation is met. Note that the features appear slightly broadened for $M_1$ due to the use of an LP instead of a BP filter. Above threshold (right), the emission condenses into bright spots: at $\theta \approx \pm 42°$ for $M_5$, $\theta \approx \pm 45°$ for $M_4$, and near-normal direction for $M_1$.

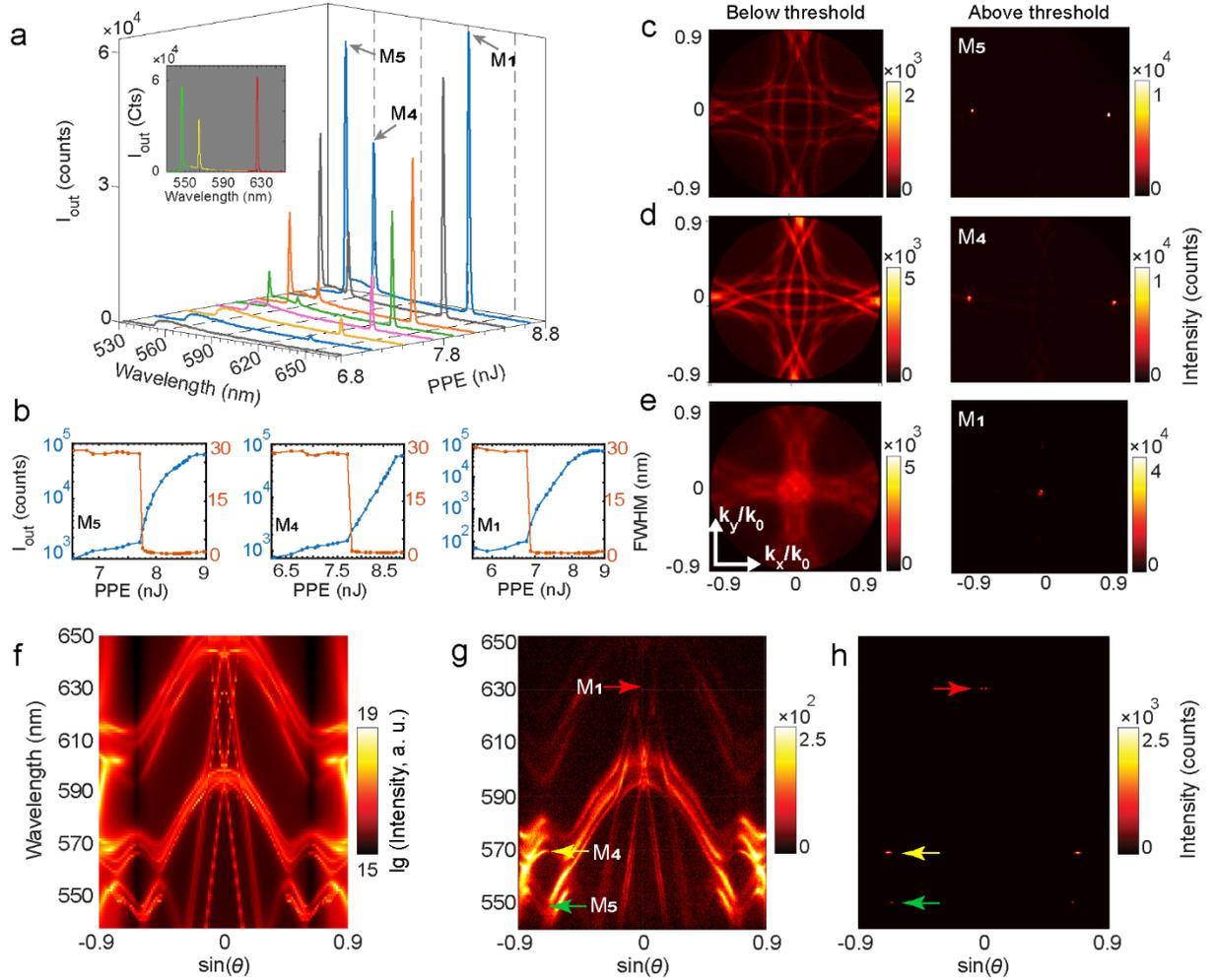

**Figure 3**. (a) Evolution of the emission spectra of LMS2 at different PPE. The inset shows the lasing spectrum (at 8.7 nJ) with discrete-frequency outputs ranging from green to red. (b) Maximum output intensity $I_{out}$ (blue) and FWHMs (orange) at different PPEs for $M_5$, $M_4$, and $M_1$. (c-e) Measured BFP images of LMS2 using BP filters centered at (c) 550 nm ($M_5$) and (d) 570 nm ($M_4$), as well as (e) an LP filter at 610 nm ($M_1$). Below the threshold (left) and with a repetition rate of 500 Hz and an integration time of 2 s to increase the signal strength, spontaneous emission occurs in the shape of broad diffraction rings from the periodic array. Above threshold (right) at PPE (1 Hz, 1 s) (c) 8 nJ, (d) 8.2 nJ, and (e) 8 nJ, lasing emission becomes dominant under $\theta \approx \pm 42°$ for $M_5$, $\theta \approx \pm 45°$ for $M_4$, and near normal direction for $M_1$.

(f) Numerically calculated angular resolved spectrum and (g, h) measured momentum-resolved spectra below (g) and above (h) threshold."

Figure 3g,h shows momentum-resolved spectra below and above threshold: below threshold, dispersive branches appear around 630 nm, folding back at Γ, consistent with an SLR band edge. The emission is weak as it lies at the tail of the Rh6G gain spectrum. Above threshold, a bright near-normal lasing peak emerges at 630 nm ($M_1$). Furthermore, two additional oblique-angle lasing modes emerge: $M_5$ (550 nm) at $\theta \approx \pm 42°$, and $M_4$ (570 nm) at $\theta \approx \pm 45°$, coinciding with the band edges observed in the below-threshold momentum spectra. The simulated angular-resolved spectra (Figure 3f) are in excellent agreement with the measured results. We investigate the origin of the modes $M_5$ and $M_4$ by calculating the normalized near-field intensity profiles at the corresponding $(\lambda, \theta)$ coordinate. From the near-field profiles (Figure S7), we identify $M_5$ as a GMR and $M_4$ as an SU8-side SLR. The latter agrees with the RA condition for P=405 nm and $n_{eff} \approx 1.57$, giving $\lambda$ =568 nm at $\theta \approx 45°$. This illustrates how period, angle, and effective index can be co-designed to direct lasing into specific emission channels. By contrast, GMRs cannot be reliably predicted by simple slab formulas in our system, since strong scattering from the embedded $TiO_2$ nanoresonators perturbs the grating-coupling condition[18].

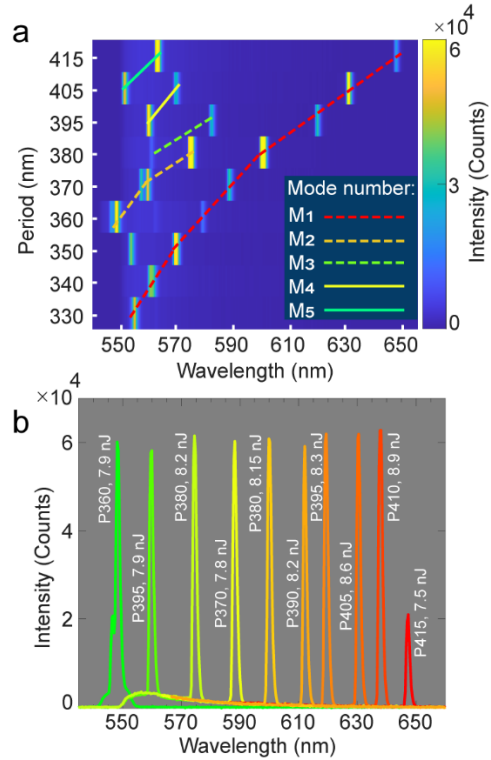

"**Figure 4**. (a) Lasing emission spectra for metasurface periods ranging from 330 nm to 415 nm. The modes contributing to lasing $M_1$–$M_5$ are shown by colored lines. (b) The spectral tunability of the lasing emission, supported by metasurfaces with different periods, is presented."

Finally, we systematically investigated metasurfaces with the same nanoresonator size and P ranging from 330 to 415 nm. Figure 4a illustrates the lasing spectra, where the main modes ($M_1$–$M_5$, indicated by colored lines) redshift with increasing period. As summarized in Figure 4b, this geometrical tunability enables tailoring of the emission wavelength(s) over a broad spectral range from 548 to 648 nm, nearly 100 nm, spanning the entire Rh6G gain bandwidth, with comparable thresholds. Notably, 548 nm (P=360 nm) and 648 nm (P=415 nm) represent the shortest and longest lasing wavelengths reported to date for Rh6G-integrated metasurfaces (Figure S8). In addition to the single, double, and triple-mode operation shown in Figure 4a, we also observe up to four simultaneous lasing peaks from a single metasurface design with a larger nanoresonator lateral dimensions (Figure S9).

In summary, we have realized a Rh6G-integrated $TiO_2$ metasurface laser that combines several complementary resonance types—GMRs, SLRs, and quasi-BICs—within a single device. The deliberate use of symmetry breaking and lattice design allows for tailoring a set of high-Q resonances with strong mode-gain overlap, enabling multifrequency lasing throughout 548–648 nm, including near the spectral limits of Rh6G. Outcoupling via second-order Bragg and Rayleigh condition allows for tailoring the direction of the laser beam, including normal and oblique directions, with consistent thresholds around ∼7 nJ PPE (35.7 µJ/cm$^2$). Importantly, our approach demonstrates not only broadband coverage but also robust multimode operation, with up to four lasing peaks achievable within a single metasurface. Looking ahead, this work offers a versatile route to engineer spectral and modal content as well as the beaming direction in compact nanolasers, with potential applications in integrated photonics, multiplexed sensing, optical communications, and advanced display technologies.[38-41]

## ASSOCIATED CONTENT

**Author Contributions**

A.B. and A.V. developed the design process. A.B. performed numerical simulations and optical characterization. K.T. fabricated the samples. M. Y. performed SEM measurements. A.B. wrote the first draft of the manuscript. T.P. and I.S. supervised the project and provided technical guidance. A.B. and I.S. revised the manuscript.

**Notes**

The authors declare no competing financial interest.

## ACKNOWLEDGMENT

We thank B. Narantsatsralt and D. Repp for valuable discussions. We thank M. Steinert for performing the FIB cut. This work was funded by the Deutsche Forschungsgemeinschaft (DFG, German Research Foundation) through the International Research Training Group (IRTG) 2675 "Meta-ACTIVE", project number 437527638, and through the Emmy Noether Program (STA 1426/2-1).

## ABBREVIATIONS

Q-factor, quality-factor; GMR, guided-mode resonance; SLR, surface lattice resonance; RA, Rayleigh anomaly; BIC, bound states in the continuum; $TiO_2$, titanium dioxide; TE, transverse-electric; $SiO_2$, silicon dioxide; Rh6G, Rhodamine 6G; P, lattice period; PPE, pump pulse energy; LMS, lasing metasurface; MQ, magnetic quadrupole; TM, transverse-magnetic; MD, magnetic dipole; EBL, electron beam lithography; RIE, reactive ion etching; SI, supporting information; SEM, scanning electron microscopy; FIB, focused-ion-beam; FWHM, full-width at half-maximum; BFP, back-focal plane; BP, band-pass; LP, long-pass.